# Properties of Elementary Fermions of the Standard Model deduced from Linear Canonical Transformations

**Ravo Tokiniaina Ranaivoson[1] , Raoelina Andriambololona[2], Hanitriarivo Rakotoson[3]**

*tokhiniaina@gmail.com[1], tokiniainaravor13@gmail.com[1], raoelina.andriambololona@gmail.com[2], jacquelineraoelina@hotmail.com[2], raoelinasp@yahoo.fr[2], infotsara@gmail.com[3]*

*Theoretical Physics and Computer Science Department*

*Institut National des Sciences et Techniques Nucléaires ( INSTN- Madagascar)*

BP 3907 Antananarivo 101 –Madagascar*, instn@moov.mg*

*Abstract*: This paper is a continuation of our works concerning Linear Canonical Transformations (LCT) and Phase Space Representation of Quantum Theory. The purpose is to study the spinorial representation of some particular LCT called Isodispersion LCT (ILCT) and to deduce a relation between them and the elementary fermions of the Standard Model of Particle Physics. After giving the definition of ILCT for the case of a general pseudo-Euclidean space and constructing their spinorial representation, we consider the particular case of a pentadimensional space with signature (1, 4). We then deduce a classification of quarks, leptons and their antiparticles according to the values of electric charge, weak hypercharge, weak isospin and colors after the introduction of appropriate operators defined from the generators of the Clifford Algebra corresponding to the ILCT spinorial representation. It is established that the electric charge is the sum of four terms, the weak hypercharge of five terms and the weak isospin of two terms. Existence of sterile neutrinos is suggested and the possibility of describing a fermions generation with a single field is discussed.

**1-Introduction**

In our previous works [1-8], we have performed studies concerning phase space representation of quantum theory and Linear Canonical Transformations (LCTs). One of the results obtained has been the construction of a spinorial representation of the LCTs [6, 8]. In the present works, our purpose is to establish the relation between the LCTs and the elementary fermions of the Standard Model of Particle Physics [9-14] i.e. quarks and leptons.

First, we define the Isodispersion Linear Canonical Transformations (ILCTs) and their spinorial representation in the case of the $N$-dimensional pseudo-euclidian space with signature ( $N_+, N_-$ ).Then, we apply this study to a pentadimensional pseudo- euclidian space, signature ( 1, 4). We find that it permits to describe the properties of the fermions belonging to a single generation of the Standard Model and their antiparticles. This pentadimensional space may be considered as the ordinary Minkowski space corresponding to relativistic spacetime, signature (1,3), with an additional fifth dimension. It may be expected that this fifth dimension is related to mass. The choice of the five dimensions is related to the fact that the total numbers of particles and antiparticles for a single generation is equal to 32 (table 1). The introduction of a right-handed (sterile) neutrino and its antiparticle is suggested by our study to complete the list of 32 particles and antiparticles.

Our method may be considered as an extension of the approach developed by Zenczykowski for nonrelativistic phase space [15-16]. He obtains that the weak hypercharge is the sum of three terms. In our case, it is the sum of five terms (see relation 3.5).

As in our previous work, we use for tensorial and matricial calculations the notations introduced in the reference [17].

## 2-Spinorial representation of Isodispersion Linear Canonical Transformations

### 2.1 Isodispersion Linear Canonical Transformations (ILCTs)

We have established [5, 6, 8], that LCTs corresponding to a $N$- dimension pseudo-euclidian space $E_N$ with signature ($N_+, N_-$) can be defined as the linear transformations which leave invariant the canonical commutation relation defining coordinate and momentum operators. They can be described with the elements of the pseudo-symplectic group $Sp(2N_+, 2N_-)$.
In the reference [6], we have introduced the following parameterization for an LCT

$$(\boldsymbol{p}' \quad \boldsymbol{x}') = (\boldsymbol{p} \quad \boldsymbol{x}) e^{\begin{pmatrix} \lambda+\mu & \varphi-\theta \\ \varphi+\theta & \lambda-\mu \end{pmatrix}} \qquad (2.1)$$

in which $\boldsymbol{p}$ and $\boldsymbol{x}$ are the $1 \times N$ matrices which admit as elements the components $\boldsymbol{p}_\mu$ and $\boldsymbol{x}_\mu$ ($\mu = 0, \ldots, N-1$) of the reduced momentum and coordinate operators. $\lambda, \mu, \varphi, \theta$ are $N \times N$ matrices verifying the relations

$$\begin{cases} \theta^T = \eta^T \theta \eta = \eta \theta \eta \\ \varphi^T = \eta^T \varphi \eta = \eta \varphi \eta \\ \mu^T = \eta^T \mu \eta = \eta \mu \eta \\ \lambda^T = -\eta^T \lambda \eta = -\eta \lambda \eta \\ Spur(\lambda) = 0 \end{cases} \qquad (2.2)$$

$\eta$ being the $N \times N$ matrices admitting as elements the components $\eta_{\mu\nu}$ of the symmetric bilinear form defining the inner product on the pseudo-euclidian space $E_N$.

$$\eta_{\mu\nu} = \begin{cases} 1 & if\ \mu = \nu = 0, 1, \ldots, N_+ - 1 \\ -1 & if\ \mu = \nu = N_+, N_+ + 1, \ldots, N - 1 \\ 0 & if\ \mu \neq \nu \end{cases} \qquad (2.3)$$

The infinitesimal form of the LCT (2.1) is

$$(\boldsymbol{p}' \quad \boldsymbol{x}') = (\boldsymbol{p} \quad \boldsymbol{x})[1 + \begin{pmatrix} \lambda+\mu & \varphi-\theta \\ \varphi+\theta & \lambda-\mu \end{pmatrix}] \qquad (2.4)$$

For $\mu = 0$ and $\varphi = 0$, we have $\begin{pmatrix} \lambda+\mu & \varphi-\theta \\ \varphi+\theta & \lambda-\mu \end{pmatrix} = \begin{pmatrix} \lambda & -\theta \\ \theta & \lambda \end{pmatrix}$

It can be deduced from the relations (2.2) that one has the relations

$$\begin{cases} \begin{pmatrix} \lambda & -\theta \\ \theta & \lambda \end{pmatrix}^T = -\begin{pmatrix} \eta & 0 \\ 0 & \eta \end{pmatrix}\begin{pmatrix} \lambda & -\theta \\ \theta & \lambda \end{pmatrix}\begin{pmatrix} \eta & 0 \\ 0 & \eta \end{pmatrix} \\ Spur\begin{pmatrix} \lambda & -\theta \\ \theta & \lambda \end{pmatrix} = 0 \end{cases} \quad (2.5)$$

The relations in (2.5) means that $\begin{pmatrix} \lambda & -\theta \\ \theta & \lambda \end{pmatrix}$ is an element of the Lie algebra $\mathfrak{so}(2N_+, 2N_-)$ of the special orthogonal group $SO(2N_+, 2N_-)$ i.e. the matrix $e^{\begin{pmatrix} \lambda & -\theta \\ \theta & \lambda \end{pmatrix}}$ belongs to $SO(2N_+, 2N_-)$

$$\begin{cases} [e^{\begin{pmatrix} \lambda & -\theta \\ \theta & \lambda \end{pmatrix}}]^T \begin{pmatrix} \eta & 0 \\ 0 & \eta \end{pmatrix} e^{\begin{pmatrix} \lambda & -\theta \\ \theta & \lambda \end{pmatrix}} = \begin{pmatrix} \eta & 0 \\ 0 & \eta \end{pmatrix} \\ Det[e^{\begin{pmatrix} \lambda & -\theta \\ \theta & \lambda \end{pmatrix}}] = 1 \end{cases} \quad (2.6)$$

The relations (2.1), (2.5) and (2.6) imply that the LCT corresponding to $e^{\begin{pmatrix} \lambda & -\theta \\ \theta & \lambda \end{pmatrix}}$ leaves invariant the operator

$$\beth^+ = \eta^{\mu\nu}\beth^+_{\mu\nu} = \frac{1}{4}\eta^{\mu\nu}(\boldsymbol{p}_\mu \boldsymbol{p}_\nu + \boldsymbol{x}_\mu \boldsymbol{x}_\nu)$$

The $\beth^+_{\mu\nu}$ is the reduced dispersion-codispersion operators introduced in the reference [5]. Explicitly, we have

$$(\boldsymbol{p}' \quad \boldsymbol{x}') = (\boldsymbol{p} \quad \boldsymbol{x})e^{\begin{pmatrix} \lambda & -\theta \\ \theta & \lambda \end{pmatrix}} \Leftrightarrow \beth'^+ = \beth^+ \quad (2.7)$$

The relation (2.7) can be checked using the relations (2.6) and the relations

$$\beth^+ = \eta^{\mu\nu}\beth^+_{\mu\nu} = \frac{1}{4}\eta^{\mu\nu}(\boldsymbol{p}_\mu \boldsymbol{p}_\nu + \boldsymbol{x}_\mu \boldsymbol{x}_\nu) = (\boldsymbol{p} \quad \boldsymbol{x})\begin{pmatrix} \eta & 0 \\ 0 & \eta \end{pmatrix}(\boldsymbol{p} \quad \boldsymbol{x})^T \quad (2.8)$$

We define the LCTs corresponding to a matrix of the form $e^{\begin{pmatrix} \lambda & -\theta \\ \theta & \lambda \end{pmatrix}} \in SO(2N_+, 2N_-)$ as the Isodispersion Linear Canonical Transformations (ILCTs). This definition is in line with the study performed for the case of one dimensional quantum mechanics in our work [3]. As we show in the reference [8], for $\theta = 0$ ($\lambda \neq 0$) an ILCT is a Lorentz-like transformation. For the case $\lambda = 0$ ($\theta \neq 0$), we have fractional Fourier- like transformations.

## 2.2 Spinorial representation

As an ILCT is defined with an element $e^{\begin{pmatrix} \lambda & -\theta \\ \theta & \lambda \end{pmatrix}}$ of the special pseudo-orthogonal group $SO(2N_+, 2N_-)$, it can be also represented spinoriarly with an element $\boldsymbol{S}$ of the spin group $Spin(2N_+, 2N_-)$ which is the double cover of $SO(2N_+, 2N_-)$. To construct this spinorial representation, we introduce, like in [6] the operator

$$\mathbb{P} = \alpha^\mu \boldsymbol{p}_\mu + \beta^\mu \boldsymbol{x}_\mu \quad (2.9)$$

in which $\alpha^\mu$ and $\beta^\mu$ are the generators of the Clifford algebra $\mathbb{C}(2N_+, 2N_-)$ i.e. they verify the following anticommutation relations :

$$\begin{cases} \alpha^\mu \alpha^\nu + \alpha^\nu \alpha^\mu = 2\eta^{\mu\nu} \\ \beta^\mu \beta^\nu + \beta^\nu \beta^\mu = 2\eta^{\mu\nu} \\ \alpha^\mu \beta^\nu + \beta^\nu \alpha^\mu = 0 \end{cases} \quad (2.10)$$

The spinorial representation can be defined explicitly by a mapping $\varrho$, from $SO(2N_+, 2N_-)$ to $Spin(2N_+, 2N_-)$, which associates to $e^{\begin{pmatrix} \lambda & -\theta \\ \theta & \lambda \end{pmatrix}}$ an element $\mathcal{S}$ of $Spin(2N_+, 2N_-)$

$$\mathcal{S} = \varrho(e^{\begin{pmatrix} \lambda & -\theta \\ \theta & \lambda \end{pmatrix}}) \Leftrightarrow \begin{cases} (\mathbb{p}' \quad \mathbb{x}') = (\mathbb{p} \quad \mathbb{x}) e^{\begin{pmatrix} \lambda & -\theta \\ \theta & \lambda \end{pmatrix}} \\ \mathbb{p}' = \mathcal{S}\mathbb{p}\mathcal{S}^{-1} \end{cases} \quad (2.11)$$

$$\mathcal{S} = e^{[\frac{1}{4}(\eta_{\mu\rho}\lambda_\nu^\rho + \eta_{\nu\rho}\lambda_\mu^\rho)(\alpha^\mu \alpha^\nu + \beta^\mu \beta^\nu) + \frac{1}{2}\eta_{\mu\rho}\theta_\nu^\rho \alpha^\mu \beta^\nu]} \quad (2.12)$$

The operator $\mathcal{S}$ acts on the elements $\vec{\psi}$ of a spinor space $\mathbb{S}$

$$\vec{\psi}' = \mathcal{S}\vec{\psi} \Leftrightarrow \psi'^a = \mathcal{S}_b^a \psi^b \quad (2.13)$$

The couple $(\mathbb{S}, \varrho)$ define the spinorial representation of an ILCT, $\mathbb{S}$ being the representation space and $\varrho$ the group morphism.

### 3-Deduction of the properties of elementary fermions

We consider the particular case of a pentadimensional pseudo-Euclidian space with signature (1, 4). The Clifford algebra and spin group corresponding to the spinorial representation of ILCTs are then respectively $\mathfrak{C}$ (2,8) and $Spin$ (2,8). From the generators $\alpha^\mu$ and $\beta^\mu$ of the Clifford algebra $\mathfrak{C}$ (2,8), we define the operators

$$\begin{cases} \mathcal{Y}^0 = \frac{i}{4}[\alpha^0, \beta^0] = \frac{1}{2}i\alpha^0\beta^0 \\ \mathcal{Y}^1 = \frac{i}{6}[\alpha^1, \beta^1] = \frac{1}{3}i\alpha^1\beta^1 \\ \mathcal{Y}^2 = \frac{i}{6}[\alpha^2, \beta^2] = \frac{1}{3}i\alpha^2\beta^2 \\ \mathcal{Y}^3 = \frac{i}{6}[\alpha^3, \beta^3] = \frac{1}{3}i\alpha^3\beta^3 \\ \mathcal{Y}^4 = \frac{i}{4}[\alpha^4, \beta^4] = \frac{1}{2}i\alpha^4\beta^4 \end{cases} \quad (3.1)$$

The operators $\mathcal{Y}^\mu$ satisfy the following properties

$$\begin{cases} (\mathcal{Y}^0)^2 = (\mathcal{Y}^4)^2 = \frac{1}{4} \quad (\mathcal{Y}^1)^2 = (\mathcal{Y}^2)^2 = (\mathcal{Y}^3)^2 = \frac{1}{9} \\ [\mathcal{Y}^\mu, \mathcal{Y}^\nu] = \mathcal{Y}^\mu \mathcal{Y}^\nu - \mathcal{Y}^\nu \mathcal{Y}^\mu = 0 \end{cases} \quad (3.2)$$

- The eigenvalues of $\mathcal{Y}^0$ and $\mathcal{Y}^4$ are $-\frac{1}{2}$ and $\frac{1}{2}$
- The eigenvalues of $\mathcal{Y}^1, \mathcal{Y}^2, \mathcal{Y}^3$ are $-\frac{1}{3}$ and $\frac{1}{3}$
- These operators commute and are simultaneously diagonalizable

These results can be directly and easily checked if we choose appropriate matrices representation of the $\alpha^\mu$ and $\beta^\mu$. Let us choose

$$\begin{cases} \alpha^0 = \sigma^1 \otimes \sigma^0 \otimes \sigma^0 \otimes \sigma^0 \otimes \sigma^0 \\ \alpha^1 = i\sigma^3 \otimes \sigma^1 \otimes \sigma^0 \otimes \sigma^0 \otimes \sigma^0 \\ \alpha^2 = i\sigma^3 \otimes \sigma^3 \otimes \sigma^1 \otimes \sigma^0 \otimes \sigma^0 \\ \alpha^3 = i\sigma^3 \otimes \sigma^3 \otimes \sigma^3 \otimes \sigma^1 \otimes \sigma^0 \\ \alpha^4 = i\sigma^3 \otimes \sigma^3 \otimes \sigma^3 \otimes \sigma^3 \otimes \sigma^1 \\ \beta^0 = \sigma^2 \otimes \sigma^0 \otimes \sigma^0 \otimes \sigma^0 \otimes \sigma^0 \\ \beta^1 = -i\sigma^3 \otimes \sigma^2 \otimes \sigma^0 \otimes \sigma^0 \otimes \sigma^0 \\ \beta^2 = -i\sigma^3 \otimes \sigma^3 \otimes \sigma^2 \otimes \sigma^0 \otimes \sigma^0 \\ \beta^3 = -i\sigma^3 \otimes \sigma^3 \otimes \sigma^3 \otimes \sigma^2 \otimes \sigma^0 \\ \beta^4 = -i\sigma^3 \otimes \sigma^3 \otimes \sigma^3 \otimes \sigma^3 \otimes \sigma^2 \end{cases} \quad (3.3)$$

in which $\sigma^0$ is the $2 \times 2$ identity matrix and $\sigma^1, \sigma^2, \sigma^3$ are the Pauli matrices

$$\sigma^0 = \begin{pmatrix} 1 & 0 \\ 0 & 1 \end{pmatrix} \quad \sigma^1 = \begin{pmatrix} 0 & 1 \\ 1 & 0 \end{pmatrix} \quad \sigma^2 = \begin{pmatrix} 0 & -i \\ i & 0 \end{pmatrix} \quad \sigma^3 = \begin{pmatrix} 1 & 0 \\ 0 & -1 \end{pmatrix}$$

The operators $\mathcal{Y}^\mu$ are represented by the following diagonal matrices

$$\begin{cases} \mathcal{Y}^0 = -\frac{1}{2}\sigma^3 \otimes \sigma^0 \otimes \sigma^0 \otimes \sigma^0 \otimes \sigma^0 \\ \mathcal{Y}^1 = -\frac{1}{3}\sigma^0 \otimes \sigma^3 \otimes \sigma^0 \otimes \sigma^0 \otimes \sigma^0 \\ \mathcal{Y}^2 = -\frac{1}{3}\sigma^0 \otimes \sigma^0 \otimes \sigma^3 \otimes \sigma^0 \otimes \sigma^0 \\ \mathcal{Y}^3 = -\frac{1}{3}\sigma^0 \otimes \sigma^0 \otimes \sigma^0 \otimes \sigma^3 \otimes \sigma^0 \\ \mathcal{Y}^4 = -\frac{1}{2}\sigma^0 \otimes \sigma^0 \otimes \sigma^0 \otimes \sigma^0 \otimes \sigma^3 \end{cases} \quad (3.4)$$

Let us now define the following operators

$$\begin{cases} I_3 = \frac{1}{2}\mathcal{Y}^0 - \frac{1}{2}\mathcal{Y}^4 \\ Y_W = \mathcal{Y}^0 + \mathcal{Y}^1 + \mathcal{Y}^2 + \mathcal{Y}^3 + \mathcal{Y}^4 \\ Q = \mathcal{Y}^0 + \frac{1}{2}\mathcal{Y}^1 + \frac{1}{2}\mathcal{Y}^2 + \frac{1}{2}\mathcal{Y}^3 \end{cases} \quad (3.5)$$

As it is seen in the table 1, these three operators correspond respectively to the electric charge, the weak hypercharge and weak isospin of quarks and leptons belonging to a fermions generation of the standard model and their antiparticles. It may be checked easily that we have the well-known relation

$$Q = I_3 + \frac{Y_W}{2}$$

The table 1 corresponds to a single fermions generation. The example of the first generation is considered: $e_L$ is the left-handed negaton and $\overline{e_L}$ is its antiparticle. $e_R$ is the right-handed negaton and $\overline{e_R}$ its antiparticle, $\nu_L$ and $\overline{\nu_L}$ are respectively the left-handed neutrino and its antiparticle. As it is well known, there is no right-handed (sterile) neutrino, $\nu_R$, in the Standard Model but their existence is suggested by this table 1.

The up and down quarks are respectively denoted $u$ and $d$ and their antiparticles $\overline{u}$ and $\overline{d}$. The lower script refers to chirality (R for right-handed and L for left-handed) and the upper script are colors (blue, green or red). The existence of the three possible quarks colors is described by the combinations of the eigenvalues of the operators $\mathcal{Y}^1, \mathcal{Y}^2$ and $\mathcal{Y}^3$.

| N° | $\mathcal{Y}^0$ | $\mathcal{Y}^1$ | $\mathcal{Y}^2$ | $\mathcal{Y}^3$ | $\mathcal{Y}^4$ | $I_3$ | $Y_W$ | $Q$ | Particle |
|---|---|---|---|---|---|---|---|---|---|
| 1 | $-1/2$ | $-1/3$ | $-1/3$ | $-1/3$ | $-1/2$ | 0 | $-2$ | $-1$ | $e_R$ |
| 2 | $-1/2$ | $-1/3$ | $-1/3$ | $-1/3$ | $1/2$ | $-1/2$ | $-1$ | $-1$ | $e_L$ |
| 3 | $-1/2$ | $-1/3$ | $-1/3$ | $1/3$ | $-1/2$ | 0 | $-4/3$ | $-2/3$ | $\overline{u}_R^{blue}$ |
| 4 | $-1/2$ | $-1/3$ | $-1/3$ | $1/3$ | $1/2$ | $-1/2$ | $-1/3$ | $-2/3$ | $\overline{u}_L^{blue}$ |
| 5 | $-1/2$ | $-1/3$ | $1/3$ | $-1/3$ | $-1/2$ | 0 | $-4/3$ | $-2/3$ | $\overline{u}_R^{green}$ |
| 6 | $-1/2$ | $-1/3$ | $1/3$ | $-1/3$ | $1/2$ | $-1/2$ | $-1/3$ | $-2/3$ | $\overline{u}_L^{green}$ |
| 7 | $-1/2$ | $-1/3$ | $1/3$ | $1/3$ | $-1/2$ | 0 | $-2/3$ | $-1/3$ | $d_R^{red}$ |
| 8 | $-1/2$ | $-1/3$ | $1/3$ | $1/3$ | $1/2$ | $-1/2$ | $1/3$ | $-1/3$ | $d_L^{red}$ |
| 9 | $-1/2$ | $1/3$ | $-1/3$ | $-1/3$ | $-1/2$ | 0 | $-4/3$ | $-2/3$ | $\overline{u}_R^{red}$ |
| 10 | $-1/2$ | $1/3$ | $-1/3$ | $-1/3$ | $1/2$ | $-1/2$ | $-1/3$ | $-2/3$ | $\overline{u}_L^{red}$ |
| 11 | $-1/2$ | $1/3$ | $-1/3$ | $1/3$ | $-1/2$ | 0 | $-2/3$ | $-1/3$ | $d_R^{green}$ |
| 12 | $-1/2$ | $1/3$ | $-1/3$ | $1/3$ | $1/2$ | $-1/2$ | $1/3$ | $-1/3$ | $d_L^{green}$ |
| 13 | $-1/2$ | $1/3$ | $1/3$ | $-1/3$ | $-1/2$ | 0 | $-2/3$ | $-1/3$ | $d_R^{blue}$ |
| 14 | $-1/2$ | $1/3$ | $1/3$ | $-1/3$ | $1/2$ | $-1/2$ | $1/3$ | $-1/3$ | $d_L^{blue}$ |
| 15 | $-1/2$ | $1/3$ | $1/3$ | $1/3$ | $-1/2$ | 0 | 0 | 0 | $\overline{\nu}_R$ |
| 16 | $-1/2$ | $1/3$ | $1/3$ | $1/3$ | $1/2$ | $-1/2$ | 1 | 0 | $\overline{\nu}_L$ |
| 17 | $1/2$ | $-1/3$ | $-1/3$ | $-1/3$ | $-1/2$ | $1/2$ | $-1$ | 0 | $\nu_L$ |
| 18 | $1/2$ | $-1/3$ | $-1/3$ | $-1/3$ | $1/2$ | 0 | 0 | 0 | $\nu_R$ |
| 19 | $1/2$ | $-1/3$ | $-1/3$ | $1/3$ | $-1/2$ | $1/2$ | $-1/3$ | $1/3$ | $\overline{d}_L^{blue}$ |
| 20 | $1/2$ | $-1/3$ | $-1/3$ | $1/3$ | $1/2$ | 0 | $2/3$ | $1/3$ | $\overline{d}_R^{blue}$ |
| 21 | $1/2$ | $-1/3$ | $1/3$ | $-1/3$ | $-1/2$ | $1/2$ | $-1/3$ | $1/3$ | $\overline{d}_L^{gree}$ |
| 22 | $1/2$ | $-1/3$ | $1/3$ | $-1/3$ | $1/2$ | 0 | $2/3$ | $1/3$ | $\overline{d}_R^{green}$ |
| 23 | $1/2$ | $-1/3$ | $1/3$ | $1/3$ | $-1/2$ | $1/2$ | $1/3$ | $2/3$ | $u_L^{red}$ |
| 24 | $1/2$ | $-1/3$ | $1/3$ | $1/3$ | $1/2$ | 0 | $4/3$ | $2/3$ | $u_R^{red}$ |
| 25 | $1/2$ | $1/3$ | $-1/3$ | $-1/3$ | $-1/2$ | $1/2$ | $1/3$ | $1/3$ | $\overline{d}_L^{red}$ |
| 26 | $1/2$ | $1/3$ | $-1/3$ | $-1/3$ | $1/2$ | 0 | $2/3$ | $1/3$ | $\overline{d}_R^{red}$ |
| 27 | $1/2$ | $1/3$ | $-1/3$ | $1/3$ | $-1/2$ | $1/2$ | $1/3$ | $2/3$ | $u_L^{green}$ |
| 28 | $1/2$ | $1/3$ | $-1/3$ | $1/3$ | $1/2$ | 0 | $4/3$ | $2/3$ | $u_R^{green}$ |
| 29 | $1/2$ | $1/3$ | $1/3$ | $-1/3$ | $-1/2$ | $1/2$ | $1/3$ | $2/3$ | $u_L^{blue}$ |
| 30 | $1/2$ | $1/3$ | $1/3$ | $-1/3$ | $1/2$ | 0 | $4/3$ | $2/3$ | $u_R^{blue}$ |
| 31 | $1/2$ | $1/3$ | $1/3$ | $1/3$ | $-1/2$ | $1/2$ | 1 | 1 | $\overline{e_L}$ |
| 32 | $1/2$ | $1/3$ | $1/3$ | $1/3$ | $1/2$ | 0 | 2 | 1 | $\overline{e_R}$ |

**Table 1:** *Classification of quarks, leptons and their antiparticles according to the eigenvalues of the operators $\mathcal{Y}^0, \mathcal{Y}^1, \mathcal{Y}^2, \mathcal{Y}^3, \mathcal{Y}^4, I_3, Y_W$ and $Q$*

## 4-Description of a fermions generation with a single field

The table 1 suggests the possibility of describing a fermions generation with a single spinor field $\vec{\psi}$ having 32 components. If we denote $\mathbb{S}$ the spinor space to which belongs the spinor $\vec{\psi}$, we may write:

$$\mathbb{S} = \mathbb{S}_e \oplus \mathbb{S}_u \oplus \mathbb{S}_v \oplus \mathbb{S}_d \tag{4.1}$$

in which: $\mathbb{S}_e$ is the subspace of $\mathbb{S}$ composed by the electron-type fields, $\mathbb{S}_u$ the subspace of up quark-type fields, $\mathbb{S}_v$ the subspace of neutrino-type fields and $\mathbb{S}_d$ the subspace of quark down-type fields. Explicitly, we have for a general field $\vec{\psi}$ itself

$$\vec{\psi} = \psi^a \vec{\zeta}_a = \vec{\psi}_e + \vec{\psi}_u + \vec{\psi}_v + \vec{\psi}_d = (\Pi_e + \Pi_u + \Pi_v + \Pi_d)\vec{\psi} \tag{4.2}$$

in which $\{\vec{\zeta}_a\}, a = 1$ to 32, is a basis of $\mathbb{S}$ constituted by the eigenspinors $\vec{\zeta}_a$ of the operators $\mathcal{Y}^0, \mathcal{Y}^1, \mathcal{Y}^2, \mathcal{Y}^3, \mathcal{Y}^4, I_3, Y_W$ and $Q$. $\vec{\psi}_e, \vec{\psi}_u, \vec{\psi}_v$ and $\vec{\psi}_d$ are respectively electron-type field ($\vec{\psi}_e \in \mathbb{S}_e$), up quark-type field ($\vec{\psi}_u \in \mathbb{S}_u$), neutrino-type field and ($\vec{\psi}_v \in \mathbb{S}_v$) and down quark-type field ($\vec{\psi}_d \in \mathbb{S}_d$). $\Pi_e, \Pi_u, \Pi_v$ and $\Pi_d$ are the projection operators corresponding to the subspaces $\mathbb{S}_e, \mathbb{S}_u, \mathbb{S}_v$ and $\mathbb{S}_d$. These projection operators can be expressed as polynomials of $\mathcal{Y}^0, \mathcal{Y}^1, \mathcal{Y}^2$ and $\mathcal{Y}^3$. Following the table 1, we have

- For the electron-type field

$$\vec{\psi}_e = \Pi_e \vec{\psi} = \psi^1 \vec{\zeta}_1 + \psi^2 \vec{\zeta}_2 + \psi^{31} \vec{\zeta}_{31} + \psi^{32} \vec{\zeta}_{32}$$

$$\Pi_e = \frac{(1 - 2\mathcal{Y}^0)(1 - 3\mathcal{Y}^1)(1 - 3\mathcal{Y}^2)(1 - 3\mathcal{Y}^3)}{16} + \frac{(1 + 2\mathcal{Y}^0)(1 + 3\mathcal{Y}^1)(1 + 3\mathcal{Y}^2)(1 + 3\mathcal{Y}^3)}{16}$$

- Up quark-type field

$$\vec{\psi}_u = \Pi_u \vec{\psi} = \psi^3 \vec{\zeta}_3 + \psi^4 \vec{\zeta}_4 + \psi^5 \vec{\zeta}_5 + \psi^6 \vec{\zeta}_6 + \psi^9 \vec{\zeta}_9 + \psi^{10} \vec{\zeta}_{10}$$

$$+ \psi^{23} \vec{\zeta}_{23}, \psi^{24} \vec{\zeta}_{24} + \psi^{27} \vec{\zeta}_{27} + \psi^{28} \vec{\zeta}_{28} + \psi^{29} \vec{\zeta}_{29} + \psi^{30} \vec{\zeta}_{30}$$

$$\Pi_u = \frac{(1 - 2\mathcal{Y}^0)(1 - 3\mathcal{Y}^1)(1 - 3\mathcal{Y}^2)(1 + 3\mathcal{Y}^3)}{16} + \frac{(1 - 2\mathcal{Y}^0)(1 - 3\mathcal{Y}^1)(1 + 3\mathcal{Y}^2)(1 - 3\mathcal{Y}^3)}{16}$$

$$+ \frac{(1 - 2\mathcal{Y}^0)(1 + 3\mathcal{Y}^1)(1 - 3\mathcal{Y}^2)(1 - 3\mathcal{Y}^3)}{16} + \frac{(1 + 2\mathcal{Y}^0)(1 - 3\mathcal{Y}^1)(1 + 3\mathcal{Y}^2)(1 + 3\mathcal{Y}^3)}{16}$$

$$+ \frac{(1 + 2\mathcal{Y}^0)(1 + 3\mathcal{Y}^1)(1 - 3\mathcal{Y}^2)(1 + 3\mathcal{Y}^3)}{16} + \frac{(1 + 2\mathcal{Y}^0)(1 + 3\mathcal{Y}^1)(1 + 3\mathcal{Y}^2)(1 - 3\mathcal{Y}^3)}{16}$$

- Neutrino-type field

$$\vec{\psi}_v = \Pi_v \vec{\psi} = \psi^{15} \vec{\zeta}_{15} + \psi^{16} \vec{\zeta}_{16} + \psi^{17} \vec{\zeta}_{17} + \psi^{18} \vec{\zeta}_{18}$$

$$\Pi_v = \frac{(1 - 2\mathcal{Y}^0)(1 + 3\mathcal{Y}^1)(1 + 3\mathcal{Y}^2)(1 + 3\mathcal{Y}^3)}{16} + \frac{(1 + 2\mathcal{Y}^0)(1 - 3\mathcal{Y}^1)(1 - 3\mathcal{Y}^2)(1 - 3\mathcal{Y}^3)}{16}$$

- Down quark-type field

$$\vec{\psi}_d = \Pi_d \vec{\psi} = \psi^7 \vec{\zeta}_7 + \psi^8 \vec{\zeta}_8 + \psi^{11} \vec{\zeta}_{11} + \psi^{12} \vec{\zeta}_{12} + \psi^{13} \vec{\zeta}_{13} + \psi^{14} \vec{\zeta}_{14}$$
$$+ \psi^{19} \vec{\zeta}_{19}, \psi^{20} \vec{\zeta}_{20} + \psi^{21} \vec{\zeta}_{21} + \psi^{22} \vec{\zeta}_{22} + \psi^{25} \vec{\zeta}_{25} + \psi^{26} \vec{\zeta}_{26}$$

$$\Pi_d = \frac{(1-2\mathcal{Y}^0)(1-3\mathcal{Y}^1)(1+3\mathcal{Y}^2)(1+3\mathcal{Y}^3)}{16} + \frac{(1-2\mathcal{Y}^0)(1+3\mathcal{Y}^1)(1-3\mathcal{Y}^2)(1+3\mathcal{Y}^3)}{16}$$
$$+ \frac{(1-2\mathcal{Y}^0)(1+3\mathcal{Y}^1)(1+3\mathcal{Y}^2)(1-3\mathcal{Y}^3)}{16} + \frac{(1+2\mathcal{Y}^0)(1-3\mathcal{Y}^1)(1-3\mathcal{Y}^2)(1+3\mathcal{Y}^3)}{16}$$
$$+ \frac{(1+2\mathcal{Y}^0)(1-3\mathcal{Y}^1)(1+3\mathcal{Y}^2)(1-3\mathcal{Y}^3)}{16} + \frac{(1+2\mathcal{Y}^0)(1+3\mathcal{Y}^1)(1-3\mathcal{Y}^2)(1-3\mathcal{Y}^3)}{16}$$

## 5-Conclusion

The table 1 shows as expected that it is possible to deduce the properties of the elementary fermions of the Standard Model from the spinorial representation of ILCT corresponding to a pentadimensional pseudo-euclidian space with signature (1,4).

According to the relation (3.5), the values of electric charges, the weak hypercharge, the weak isospin and colors can be considered as corresponding to the eigenvalues of linear combinations of the operators $\mathcal{Y}^\mu$ defined in the relation (3.1) from the generators $\alpha^\mu$ and $\beta^\mu$ of the Clifford algebra $\mathfrak{C}$ (2,8).

This work brings some new point of views concerning the relation between spacetime, momentum, energy and particles. It shows that phase space in quantum theory, linear canonical transformations and particles are deeply linked.

The results may be exploited and extended to build a new theory of particles interactions beyond the Standard Model. It is worth pointing out that LCTs permit at the same time to describe linear mixing of spacetime and momentum-energy and the change of particles flavours.

.